# Estimating Stochastic Production Frontiers: A One-stage Multivariate Semi-Nonparametric Bayesian Concave Regression Method

José Luis Preciado Arreola[†] and Andrew L. Johnson[*†]

**Abstract**

This paper describes a method to estimate a production frontier that satisfies the axioms of monotonicity and concavity in a non-parametric Bayesian setting. An inefficiency term that allows for significant departure from prior distributional assumptions is jointly estimated in a single stage with parametric prior assumptions. We introduce heteroscedasticity into the inefficiency terms by local hyperplane-specific shrinkage hyperparameters and impose monotonicity using bound-constrained local nonlinear regression. Our minimum-of-hyperplanes estimator imposes concavity. Our Monte Carlo simulation experiments demonstrate that the frontier and efficiency estimations are competitive, economically sound, and allow for the analysis of larger datasets than existing nonparametric methods. We validate the proposed method using data from 2007-2010 for Japan's concrete industry. The results show that the efficiency levels remain relatively high over the time period.

[†] Department of Industrial and Systems Engineering, Texas A&M University

* Corresponding Author: Email: ajohnson@tamu.edu.  Address: 3131 TAMU, College Station, TX 77843. Phone: 979-458-2356.

# 1. Introduction

The estimation of flexible production, cost, and utility functions that globally satisfy certain second order shape restrictions consistent with economic theory, such as monotonicity and convexity/concavity, remains challenging (Diewert and Wales, 1987). Isotonic Regression, Constraint Weighted Bootstrapping, and Data Sharpening have been devised to impose such restrictions (Henderson and Parmeter, 2009), but no solution methods so far are able to both impose concavity and handle very large datasets. Motivated by the challenge, we are also interested in models that allow for firms to make errors in optimization and thus model firm inefficiency. Because the inefficiency is not directly measurable, we use a Stochastic Frontier Analysis (SFA) framework (Aigner et al., 1977). Recently, several non-parametric estimators that include inefficiency, such as Kumbhakar et al. (2007)'s estimator, Stochastic Nonparametric Envelopment of Data (Kuosmanen and Kortelainen, 2012), and Constraint Weighted Bootstrapping (Du, Parmeter and Racine, 2013) have been devised, but even these methods are limited by computational issues or homoscedasticity assumptions on the inefficiency term or both. To the best of our knowledge, Kumbhakar et al. (2007) is the only frontier production function estimation method that allows the inefficiency term to be heteroscedastic without additional observable variables that predict inefficiency; however, this method does not impose shape constraints.

      A survey of the relevant literature indicates that Banker and Maindiratta (1992) are the first to propose an estimator with second order shape restrictions and a composed error term; however, their maximum likelihood methods are only tractable for small instances and no applications or Monte Carlo studies of their method have been reported. Allon et al. (2005) use entropy methods, but little is known about their estimator's computational

performance for datasets of more than a few hundred observations. Least Squares approaches, such as Convex Nonparametric Least Squares (CNLS) (Kuosmanen, 2008) or the Penalized Least Squares splines approach by Wu and Sickles (2013), which allow estimation of production frontiers with minimal assumptions about the residual term require a separable and homoscedastic distribution for the inefficiency term.[1] Constraint Weighted Bootstrapping (CWB) (Du, Parmeter and Racine (2013)) can impose a vast array of derivative-based constraints, including both monotonicity and concavity. The concavity constraints in CNLS and CWB require satisfying $O(n^2)$ constraints simultaneously which is computationally difficult (Lee et al. 2013). Moreover, their use in a stochastic frontier setting requires a two-stage procedure, such as Kuosmanen and Kortelainen (2012), which does not allow a feedback structure between the frontier estimation procedure and the inefficiency estimation procedure. Thus, prior distributional assumptions about the inefficiency distribution, namely the family of distributions and homoscedasticity, are imposed throughout the frontier estimation procedure.

An alternative two-stage procedure by Simar and Zelenyuk (2011) adapts the Local Maximum Likelihood method developed by Kumbhakar et al. (2007) to shape constrained estimation by using Data Envelopment Analysis on the fitted values obtained from the Local Maximum Likelihood method. The approach is constrained by the Local Maximum Likelihood method, and scalability appears to be limited to a few hundred observations (Kumbhakar et al., 2007).

---

[1] While the homoscedastic assumption can be relaxed by using a specific parametric form on heteroscedasticity, see Kuosmanen and Kortelainen (2012), general models of heteroscedasticity are not available.

The first Bayesian SFA semiparametric estimators in the general multivariate setting proposed by Koop and Poirier (2004) and Griffin and Steel (2004) do not address the imposition of a concavity constraint on the estimations. O'Donnell and Coelli (2005) impose homogeneity, monotonicity, and convexity in inputs on a parametric multi-output, multi-input production frontier for a Panel dataset by means of a Metropolis-Hastings (M-H) random walk algorithm and restricting the Hessian matrix. While the approach is feasible in the parametric setting, a nonparametric equivalent requires numerical estimates of the production function derivatives, similar to Du et al. (2013), and significant computational effort.

When shape restrictions between the dependent variable and each regressor are imposed separately, Shively et al. (2011) estimate Bayesian shape-constrained nonparametric regressions using fixed and free-knot smoothing splines. The knots are endogenously inferred using a Reversible Jump Markov Chain Monte Carlo (RJMCMC) (Green 1995) algorithm. Unfortunately, the method can result in complex and numerous conditions, and consequently low acceptance rates within the parameter-sampling rejection algorithm. Furthermore, the RJMCMC algorithm cannot be directly extended to a general multivariate setting. Meyer, Hackstadt and Hoeting (2011), who sample from a set of basis functions for which the imposition of concavity constraints only relies on the non-negativity of their coefficients, thus reducing complexity, avoid rejection sampling or M-H methods. Nevertheless, the monotonicity and concavity constraints are still imposed only separately between each regressor and the dependent variable and a partial linear model form is assumed for the multivariate case.

The only Bayesian semi-nonparametric constrained method for a general multivariate context is the Neural Cost Function (Michaelides et al. 2015). Even though it can impose shape restrictions a priori to estimate a cost function, the method relies on an exogenous model selection criterion to select the number of intermediate variables, focuses on an average cost function rather than a frontier, and potentially has an overfitting problem due to near perfect correlation between predicted and actual costs.

Hannah and Dunson (2013) and Hannah and Dunson (2011) propose two adaptive regression-based multivariate nonparametric convex regression methods for estimating conditional means: Least-Squares based Convex Adaptive Partitioning (CAP), and a Bayesian method, Multivariate Bayesian Convex Regression (MBCR), both of which scale well in large data. Unlike CAP, the Markov Chain Monte Carlo nature of MBCR (Hannah and Dunson, 2011, henceforth H-D) allows to create extensions of the method in a modular manner and without risking its convergence guarantees. MBCR approximates a general convex multivariate regression function with the maximum value of a random collection of hyperplanes. Additions, removals, and changes of proposed hyperplanes are done through a RJMCMC algorithm. MBCR's attractive features include the block nature of its parameter updating, which causes parameter estimate autocorrelation to drop to zero in tens of iterations in most cases, the ability to span all convex multivariate functions without need for any acceptance-rejection samplers, scalability to a few thousand observations, and relaxation of the homoscedastic noise assumption.

To model inefficiency, this paper extends MBCR to an MBCR-based semi-nonparametric SFA method. Developing our estimator in the Bayesian context allows us to learn about the inefficiency distribution beyond prior assumptions and to obtain heteroscedastic firm-specific inefficiency estimates that are shrunk both to local variance parameters and to a common population value. The shape constrained frontier and the components of the error term are jointly estimated in a single stage. The proposed method,

MBCR with Inefficiency (henceforth MBCR-I) is computationally efficient and provides straightforward inference, returning the posterior distributions of the estimated parameters. These characteristics are unique among the estimators available in the literature. Specifically, MBCR-I is unique among SFA estimators in the literature by combining a one-stage framework, shape constraints on the frontier, heteroscedastic posterior distributions of inefficiency that can depart from the homoscedastic prior, a heteroscedastic error term, and computational feasibility for large datasets.

      The remainder of this paper is organized as follows. Section 2 describes the SFA model that we use to fit the observed data, the H-D's MBCR regression method, and our proposed method, MBCR-I) and its characteristics.  Section 3 presents our Monte Carlo simulations for comparing the performance of MBCR-I against Stochastic Nonparametric Envelopment of Data (StoNED), a nonparametric method to fit production frontiers, on the basis of three criteria: functional estimation accuracy, mean inefficiency estimation accuracy and estimator variability across replicates from the same DGP for several generated datasets.  Further, we discuss MBCR-I's capability to produce full-dimensional hyperplanes (Olesen and Petersen, 2003) and the dataset characteristics for which MBCR-I use is recommended. Section 4 discusses several extensions allowing MBCR-I to model time trends of the frontier shift for Panel data and datasets with contextual variables. Section 5 applies MBCR-I to estimate a production frontier for the Japanese concrete industry and analyzes the substitution rates, most productive scale size, and inefficiency estimates. Section 6 summarizes MBCR-I's features, empirical results, and strengths and limitations, and sets directions for future research.

## 2. Methodology

*2.1 Production frontier model*

We define the regression model for our semi-nonparametric estimation procedure as

$$Y = f(X)e^{\varepsilon}, \qquad (1)$$

where $Y$ represents observed output, $f(X)$ denotes the best attainable output level, given a certain input mix $X = (X_1, \ldots, X_k)'$, and $\varepsilon = v - u$ is a composite error term obtained by subtracting a non-negative, skewed random variable $u$ representing a firm's technical inefficiency from a symmetric random effect $v$, which we term *noise*, assuming a mean 0. For our estimation purpose, we use the firm-specific equation in (2) to derive our likelihood function for $v_i$.

$$\ln(Y_i) = \ln\bigl(f(X_{1i}, \ldots, X_{ki})\bigr) + v_i - u_i, \quad i = 1, \ldots, n \qquad (2)$$

For notational simplicity, we let $f_i = f(X_{1i}, \ldots, X_{ki})$ and $X_i = X_{1i}, \ldots, X_{ki}$. This allows us to describe the decreasing marginal productivity (concavity) property in terms of $\nabla f(X)$, i.e., the gradient of $f$ with respect to $X$,

$$f(X_i) \leq + f(X_j) + \nabla f(X_j)^T (X_i - X_j) \ \forall i, j. \qquad (3)$$

Given that the constraints in (3) hold, the additional constraint $\nabla f(X_i) > 0 \ \forall i$ imposes monotonicity.

*2.2 Multivariate Nonparametric Bayesian Concave Regression*

Even though equation (3) leads to a series of pairwise constraints that can be difficult to impose for even moderate datasets, H-D note that the global concavity constraint is met automatically for the class of functions encompassing the minimums of K hyperplanes. Moreover, H-D prove that the piecewise planar functions estimated by the MBCR algorithm are able to consistently estimate any continuous and concave function. Thus, following H-D, we estimate the function

$$\hat{f}(X) = \min_{k \in \{1,\ldots,K\}} \alpha_k + \beta_k^T X \quad (4)$$

to approximate the concave function $f(X)$.

The estimation procedure in H-D is based on proposing additions, relocations, and removals of hyperplanes, the coefficients of which are determined by fitting Bayesian linear regressions. The MBCR algorithm fits $K^{(t)}$ approximating hyperplanes at a given iteration $t$. Given the current $K^{(t)}$ hyperplanes, H-D partition the set of all observations into subsets $C_k, k \in \{1, \ldots, K\}$, term each subset a *basis region*, and define each one by $C_k = \left\{i : k = \arg\min_{k \in \{1,\ldots,K\}} \alpha_k + \beta_k^T X_i\right\}$. The MBCR algorithm decides the type of move based on the current status of the Markov Chain. After choosing the type of move, the procedures (see Figure 1) are:

*If a hyperplane is added:* The basis region is split, which creates two proposed basis regions. To divide each region, we consider *L* different proposal splitting knots and M different proposal search directions along each knot,[2] i.e, each region now has *LM* splitting proposals.

*If a hyperplane is relocated:* The current basis regions are kept except for minor changes due to refitting.

---

[2] The number of hyperplanes estimated is sensitive to the knot and direction selection criteria when adding a hyperplane. Hannah and Dunson (2011) create the knot and direction proposals randomly and we also implement this knot and direction proposal generation scheme.

*If a hyperplane is removed:* Basis regions are proposed for the $K^{(t)} - 1$ remaining hyperplanes.

Regardless of the move type chosen at iteration *t*, the hyperplane coefficients for each new basis region proposal are obtained after refitting. As is common in Bayesian analysis, prior distributional assumptions are placed on the parameters to be estimated by Bayesian linear regression for each basis region. Due to the difficulties resulting from the homoscedasticity assumption, H-D specify hyperplane-specific Gaussian noise distributions with variances $(\sigma_k^2)_{k=1}^K$.

*2.3 A multiplicative production frontier*

To fit the model described in (1) using (2), we assume $v_i$ follows the Gaussian mixture distribution $v_i \sim \mathrm{N}(\ln(Y_i) - \ln(\hat{f}_i) + u_i, \sigma_{[i]}^2)$, where $\sigma_{[i]}^2$ is the noise variance of the basis region that includes the $ith$ observation. Unlike H-D, which consider a conjugate Multivariate Normal-Inverse Gamma (NIG) prior for estimating the proposal distributions of the hyperplane coefficients $(\alpha_k, \beta_k)_{k=1}^K$ and the hyperplane-specific noise variances $(\sigma_k^2)_{k=1}^K$, we cannot rely on such conjugate proposal distributions for $(\alpha_k, \beta_k)_{k=1}^K$. Specifically, the logarithm operator applied to $\hat{f}_i$ prevents the Multivariate Normal distribution on $(\alpha_k, \beta_k)_{k=1}^K$ from being conjugate, given the Gaussian mixture likelihood function on $v_i$, shown in (5). We prioritize computational performance and forgo the ability to draw from the full posterior distributions of $(\alpha_k, \beta_k)_{k=1}^K$, instead estimating the hyperplane parameters by nonlinear least squares with a lower bound of 0 for all $\beta_k$'s to impose monotonicity.[3] Like Denison, Mallick and Smith (1998), who compute regression coefficients by least squares, but conduct the remaining analysis on a Bayesian framework, we conduct a Bayesian analysis of the remaining parameters to preserve the key MBCR property of the endogenous estimation of $K$. Recalling equation (4), we estimate $f_i = f(\mathbf{X}_i; \theta)$ by

$$Y_i = \hat{f}_i \cdot e^{v_i} \cdot e^{-u_i}, \qquad v_i \sim N\big(\ln(Y_i) - \ln(\hat{f}_i) + u_i, \sigma^2_{[i]}\big), \qquad u_i \sim H \qquad (5)$$

$$\beta_k > \mathbf{0}, \quad \sigma_k^2 \sim IG(a,b), \quad k = 1, \dots, K$$

$$K - 1 \sim Poisson(\lambda).$$

As mentioned, $\beta_k > \mathbf{0}$, $k = 1, \dots, K$ are necessary to impose monotonicity, whereas the concavity constraints are automatically satisfied, given the construction of the function set from which we choose $\hat{f}$. Initially, we consider different options for $H$, the distribution of the prior inefficiency terms $u_i$, the most general of which correspond to the Gamma distribution with two unknown continuous parameters, $\Gamma(P, \theta)$ as in Tsionas (2000). While Tsionas (2000) is able to estimate the shape parameter $P$, which Ritter and Simar (1997) show to be difficult in a frequentist setting unless several thousand observations are available, Tsionas obtains parameter estimates close to their true valuesonly when at least 1000 observations are available. Moreover, since Tsionas (2000) estimates $P$ and $\theta$ in a parametric regression setting, we expect that in a nonparametric setting this moderate-sample bias will be larger if $P$ is at all identifiable. In fact, our experiments with generated datasets indicate that $P$ is not identifiable in our nonparametric setting, even if a few thousand observations are available and a single input is considered.

---

[3] Given a vague prior on each $(\alpha_k, \beta_k)$, this is equivalent to the Maximum a Posteriori (MAP) estimate obtained from a Bayesian estimation; see Appendix A for a fully Bayesian version of the algorithm.

Therefore, we evaluate scenarios considering the Exponential and Half-Normal prior inefficiency distributions first presented in van den Broeck et al. (1994). Moreover, we place a Poisson prior on the number of hyperplanes, $K$. While this prior is not to be multiplied against the likelihood function in order to obtain a posterior distribution, we still need it to determine the addition, relocation, and removal probabilities at each iteration of MBCR as described by equations (6), where $c \in (0, 0.5]$ is a tunable parameter.

$$b_{K^{(t)}} = c\, min\left\{1, \frac{p(K^{(t)}+1)}{p(K^{(t)})}\right\}, \quad d_{K^{(t)}} = c\, min\left\{1, \frac{p(K^{(t)}-1)}{p(K^{(t)})}\right\}, \quad r_{K^{(t)}} = 1 - b_{K^{(t)}} - d_{K^{(t)}} \quad (6)$$

Equation (7) describes the mathematical program used to fit the hyperplanes and obtain $(\alpha_k, \beta_k)_{k=1}^{K}$, where $Y_{i[k]}$ and $X_{i[k]}$ refer to the $i$th observation in basis region $C_k$ and $n_k$ refers to the number of observations in the basis region. Due to the conjugacy of the IG prior with each $\sigma_k^2$, we can easily sample these posterior variances for each of the proposed basis regions by using (8). For all cases, we assume that $\theta$, the scale parameter of our prior inefficiency distribution, has a prior $\Gamma(w_0)$ distribution, as shown in (9). We also assume $w_0 = -1/\ln(\tau^*)$ and that $\tau^*$ is a prior estimate of the median Technical Efficiency. The posterior distributions for either the Exponential or the Half-Normal prior assumptions on $u_i$ are the Truncated Normals shown in (9a) and (9b), respectively, and $\varepsilon_i = \ln(Y_i) - \ln(x_i' \beta)$ denotes the residuals.

$$\min_{\alpha_k, \beta_k} \sum_{i=1}^{n_k} ((\ln(Y_i) + u_i - \ln(\alpha_k + \beta_k^T X_i))^2 \text{ subject to } \beta_k > 0, \quad k = 1, \ldots, K \quad (7)$$

$$\sigma_k^2 \sim IG(a_k^*, b_k^*), \quad k = 1, \ldots, K, \text{ where} \quad (8)$$

$$a_k^* = \tilde{a} + \frac{n_k}{2}, \quad b_k^* = \tilde{b} + \frac{1}{2}\left(\sum_{i=1}^{n_k}(\ln(Y_{i[k]}) + ui - \ln(\alpha_k + \beta_k^T X_{i[k]}))^2\right)$$

$$u_i | \ldots \propto \exp(-1/2\sigma_{u_i}^2 \cdot (\mu_{u_i} - u_i)), \quad u_i \geq 0, \quad i = 1, \ldots, n \text{ where} \quad (9a)$$

$$\mu_{u_i} = -(\varepsilon_i + \theta \sigma_{[i]}^2) \quad \sigma_{u_i}^2 = \sigma_{[i]}^2$$

$$u_i| \ldots \propto \exp(-1/2\sigma_{u_i}^2 \cdot (\mu_{u_i} - u_i)), \ u_i \geq 0, \ i = 1, \ldots, n \ \text{where} \tag{9b}$$

$$\mu_{u_i} = \frac{-\sigma_{0u}^2 \varepsilon_i}{\sigma_{0u}^2 + \sigma_{[i]}^2}, \ \sigma_{u_i}^2 = \frac{\sigma_{0u}^2 \sigma_{[i]}^2}{\sigma_{0u}^2 + \sigma_{[i]}^2}$$

$$\theta| \ldots \sim \Gamma(n+1, w_0 + \sum_{i=1}^n u_i) \tag{10}$$

*2.4 The Proposed MBCR-I Algorithm*

We propose an algorithm, MBCR-I, with a smoothed and a non-smooth variant. Our Metropolis-Hastings algorithm first calculates the block $(\alpha_k, \beta_k, \sigma_k^2)_{k=1}^K$ from our multiplicative error version of MBCR, then draws the block $(u_i)_{i=1}^n$ on a Gibbs step, and ends by drawing $\theta$ on another Gibbs step. After verifying that MBCR's fast convergence is around 100 iterations for the examples presented in H-D,[4] we consider a burn-in period for the Metropolis-Hastings sampling algorithm of 150 iterations which is safely beyond the needed convergence period. Then, we monitor mean squared error at iteration $t$, MSE$y$ ($t$) = $\frac{1}{n}\sum_{i=1}^n (\hat{Y}_i - Y_i)^2$, where $\hat{Y}_i = \hat{f}_i e^{-\hat{u}_i}$. We declare that the MBCR-I algorithm has reached stationarity when the running median does not change significantly and the variability across iterations is constant for at least 200 iterations.[5] Finally, as we usually obtain a few hundred draws from the sampling algorithm, we average the functional estimates across iterations to obtain a smoothed estimator, or we select a single iteration for a non-smooth estimator, resulting in two versions of the MBCR-I algorithm, henceforth MBCR-I S and MBCR-I NS. We find that the non-smooth estimator performs better in small datasets for which the inefficiency model is mis-specified, although it relies on a heuristic criterion to select the best iteration and inference is not possible, whereas the smoothed estimator performs well in all other settings, without relying on heuristics and inference is available directly from MBCR-I's output.

---

[4] The numerical results are available from the authors upon request.
[5] Under this stopping criterion, MBCR-I rarely needs more than 1000 iterations to reach stationarity.

Our iteration selection criterion for the non-smooth estimator is motivated by observing that for small sample sizes ($n < 200$), MBCR-I can overfit the data with highly flexible inefficiency terms. To prevent overfitting, we choose a stationary iteration with relatively conservative MSE$y$.[6] Finally, we note that as $n$ increases, the iteration selection criterion becomes irrelevant at $n \geq 300$ as the variability decreases across the iterations. This decrease in variability is the reason why the smoothed and non-smooth estimators are increasingly similar in $n$. We summarize the MBCR-I algorithm as follows:

0. Let $t = 1$, $K = 1$ and set $t_{Burn-in}$, draw $(u_i)_{i=1}^n$ and $\theta$ from their priors.

1. Use MBCR to get $((\alpha_k, \beta_k, \sigma_k^2)_{k=1}^K, K)^{(t)}$.

2. Draw $(u_i^{(t)})_{i=1}^n$ from (9a) or (9b), depending on the prior assumption.

3. Draw $\theta^{(t)}$ from (10).

4. If $t > t_{Burn-in}$, save $((\alpha_k, \beta_k, \sigma_k^2)_{k=1}^K, (u_i)_{i=1}^n, \theta, K)^{(t)}$ draw and compute

$MSE_y^{(t)} = \frac{1}{n}\sum_{i=1}^n (\hat{Y}_i - Y_i)^2$. Otherwise, go back to 1.

5. Stop when the cumulative median of $MSE_y$ meets the stationarity criterion. Otherwise, go back to 1.

6a. To obtain a smoothed estimator: Average $(\hat{f}_i)_{i=1}^n$ across the stationary iterations for the $f$ estimator and average mean inefficiency across the stationary iterations to obtain $\widehat{E(\bar{u})}$.

---

[6] Our simulations consider the iteration with maximum $MSE_y$ within the described subset of iterations.

6b. To obtain a non-smooth estimator: Choose an iteration according to the iteration selection criterion and return the parameters $((\alpha_k, \beta_k)_{k=1}^{K}, (u_i)_{i=1}^{n}, \theta, K)$ associated with that iteration.

*2.5 Additional computational considerations due to inefficiency modeling*

Incorporating inefficiency into the MBCR algorithm requires the following adaptations and augmentations. First, we need an efficient and robust Truncated Normal sampler for the inefficiency terms to quickly sample from extreme tails and avoid stalling. We use a MATLAB implementation of Chopin's (2011) fast truncated normal sampling algorithm by Mazet (2014), because other samplers do not achieve the degree of accuracy or posterior coverage needed to make MBCR-I computationally feasible. Second, as described in the MBCR-I algorithm, at any given iteration *t*, we sample the inefficiency draws $(u_i)_{i=1}^{n}$ as a block after computing the hyperplane coefficients and simulating the associated variances as a different parameter block. Here, the only step of the algorithm in which the model size is allowed to change is when we draw the block $((\alpha_k, \beta_k, \sigma_k^2)_{k=1}^{K}, K)$.

Nevertheless, we observe that due to the differences in the $(u_i)_{i=1}^{n}$ values from iteration *t* to iteration *t+1*, the number of hyperplanes supporting a positive number of observations can change even if the move at iteration *t+1* is only a relocation (or conversely, the number of hyperplanes supporting a positive number of observations remains the same even if the move is an addition or removal). We automatically reject such proposal distributions and we draw different $(u_i)_{i=1}^{n}$ values, given the $(\alpha_k, \beta_k, \sigma_k^2)_{k=1}^{K}$ of iteration *t*. If this rejection policy results in stalling, measured as the time taken to generate the (*t+1*)*th* draw compared to the average time to generate a draw, we restart the Markov

Chain $((\alpha_k, \beta_k)_{k=1}^{K}, (u_i)_{i=1}^{n}, \theta, K)$ at its value on a randomly chosen previous iteration. The restarting policy is analogous to H-D's chain restarting policy when the number of tries to produce the (*t*+1)*th* draw of the chain goes above a preset threshold. Third, we run a small number of warm-up iterations, i.e., 20, in our simulation scenarios. In these iterations, which are a subset of the burn-in iterations, we draw $(u_i)_{i=1}^{n}$ from their prior, and we run MBCR to get a good initial guess of *K*, as opposed to the *K* = 1 starting value chosen by H-D. Otherwise, the $(u_i)_{i=1}^{n}$ draws from the initial iterations will be heavily overestimated and complicate, or even prevent, the MBCR-I algorithm from running fluently. An alternative to the warm-up iterations is to use the multiplicative-error MBCR estimates of $((\alpha_k, \beta_k)_{K}^{k=1}, K)$ for the same dataset. Section 5 and Appendix C below illustrate our use of the latter strategy when analyzing the Japanese concrete industry.

*2.6 MBCR-I as a one-stage estimator for stochastic frontiers*

Unlike StoNED (Kuosmanen and Kortelainen, 2012) and Constraint Weighted Bootstrapping (CWB) (Du et al. 2013), the ability of MBCR-I to significantly depart from prior distributional assumptions on $u_i$ makes the method more robust against model mis-specifications for the inefficiency term. Our posteriors show that besides globally shrinking the inefficiency terms using either $\theta$ or $\sigma_{0u}^2$, we can *locally* shrink them with the $\sigma_{[i]}^2$ parameters.[7] Moreover, within the bounds established by the shrinking parameters, doing so allows each $u_i$ to have a potentially *different* posterior distribution. In Section 3, we explain how the posterior specification in (9a) allows a significantly better prediction of mean inefficiency and the production frontier when the inefficiency distribution is mis-specified. Our posterior specifications are not a source of additional inaccuracy when the prior distributions are in fact correct. Finally, the use of a one-stage framework imposes a correctly-skewed distribution of $\varepsilon_i$ at each iteration of the MBCR-I algorithm and avoids the wrong skewness issues of two-stage methods (see, for example, Almanidis and Sickles, 2012).

---

[7] Shrinking refers to the common parameter shrinkage concept in Hierarchical Regression Models (HRM), where parameters are constrained by a common distribution. For further discussion see Gelman and Hill (2006).

## 3. Monte Carlo Simulations

This section describes Monte Carlo simulations and their results comparing the performance of MBCR-I versus StoNED on the Data Generation Processes (DGPs) used in Kuosmanen and Kortelainen (2012), some of which are presented in Simar and Zelenyuk (2011).[8] These DGPs, henceforth Example 1 through 4, are based on Cobb-Douglas production functions and they explore the performance of both estimators as dimensionality, noise-to-signal ratio, and sample size vary. Example 2 is added to the DGPs in Kuosmanen and Kortelainen (2012) for completeness, as they do not include a bivariate input example. Example 4 assesses the robustness of each method against mis-specification on the prior inefficiency distribution. For all four examples we consider three noise-to-signal scenarios, $\rho_{nts} = 1,2,3$, and vary the number of observations, $n = 100, 200, 300,$ and $500$. StoNED is the only shape constrained frontier estimation method that can handle more than a few hundred observations under a multiplicative error assumption, and therefore is the most natural benchmark for comparison.

---

[8] CWB is a state-of-the-art two-stage method that can be used to estimate production frontiers. Nevertheless, its application is not straightforward for the proposed DGPs, because the CWB formulation in Du, Parmeter and Racine (2013) considers an additive error structure.

We compare the estimators based on criteria including quality and degree of variability in frontier estimates and quality of the inefficiency estimate. We measure the frontier estimation performance as $MSE\ f = \frac{1}{n}\sum_{i=1}^{n}(\hat{f}_i - f_i)^2$. As measures of degree of variability in our frontier estimates, we also report the number of replications needed to obtain stable estimates in terms of near-constant mean and standard deviation of $MSE\ f$. Instead of using $MSE\ u = \frac{1}{n}\sum_{i=1}^{n}(\hat{u}_i - u_i)^2$ to measure the accuracy of our inefficiency estimation as in Kuosmanen and Kortelainen (2012), we use $\widehat{E(\bar{u})} - E(\bar{u}) = \frac{1}{n}\sum_{i=1}^{n}\hat{u}_i - \frac{1}{n}\sum_{i=1}^{n}u_i$, the mean inefficiency prediction deviation.[9] MBCR-I is advantageous because its estimates typically are more consistent with production theory. Specifically, we report the number of observations supported by fully dimensional hyperplanes and the percentage of replicates for which StoNED has negatively skewed residuals.

For MBCR-I, we conduct a MATLAB implementation, considering an Exponential prior with parameter $\theta$. We randomly draw the prior values used for the $\theta$ parameter from ranges described in each of the examples. In Tables 1-8 we show results for both MBCR-I S and MBCR-I NS. In the case of StoNED, since multiplicative CNLS is a mathematical program with a generally nonlinear objective function its solutions are sensitive to the choice of starting point and solver. Thus, we conduct several implementations of CNLS and select the one with the lowest $MSE\ f$ for each example. Our MATLAB CNLS implementations use the built-in fmincon solver and the KNITRO solver. Our GAMS implementations uses the MINOS 5.5 solver. For every implementation, we consider different starting points, such as the (global) optimal solution of additive CNLS, single hyperplane solutions, and full vectors of zeros.

---

[9] The metric $MSE\ u$ focuses on firm specific efficiency estimates from the Jondrow et al. (1982) estimator that have been shown to be inconsistent, Greene (2008). Instead, we measure the quality of the inefficiency estimate based on the population parameter $E(\bar{u})$.

*3.1 Evaluation Based on Four Data Generation Processes*

*Example 1: Univariate Cobb-Douglas frontier with homoscedastic inefficiency terms*

We fit the univariate Cobb-Douglas frontier $Y_i = x_i^{0.5} e^{-u_i} e^{v_i}$, considering a homoscedastic DGP with $u_i \sim \text{Expo}(\mu_u = \sigma_u = 1/6)$ and $v_i \sim N(0, \sigma_v^2)$, where $\sigma_v = \rho_{nts} \sigma_u$. We randomize the nearly uniformative prior on $\theta$ across replicates by choosing $v_0 = 1$ and drawing $w_0$ uniformly on the (0.1, 0.2) range. Table 1 shows that in terms of $MSE\ f$, StoNED only outperforms MBCR-I S for a noise-to-signal ratio of 1 and less than 300 observations. MBCR-I S and StoNED perform similarly for a noise-to-signal ratio of 1 and $n \geq 300$. For $\rho_{nts} = 2,3$, MBCR-I S outperforms StoNED for all $n$.[10] In terms of the quality of efficiency estimates, all of the estimators perform similarly for a noise-to-signal ratio of 1, but the MBCR-I estimators are superior when there is a larger noise-to-signal ratio. Finally, the results for StoNED show that non-full dimensional hyperplanes support 1%-8% of the observations.

Table 2 shows that the percentage of replicates with a negatively skewed $\varepsilon_i$ distribution for StoNED is non-decreasing for the noise-to-signal ratio, $\rho_{nts}$, and non-increasing in the sample size, $n$, as expected. As explained in the previous section, this problem does not affect MBCR-I, because it is a one-stage method and automatically imposes correct skewness on the distribution of $\varepsilon_i$. For *MSE f*, both methods need a small number of replicates to reach relatively constant[11] values; StoNED needs only 10 replicates for all scenarios, whereas MBCR-I needs 20 replicates for 2 of the 12 considered scenarios We attribute the variability of MBCR-I's prediction error, which is smaller both in absolute and relative terms as quantified by the standard deviation and the coefficient of variation of *MSE f* across all replicates, as the result of non-smooth MBCR-I fitting the production frontier with a smaller number of hyperplanes than StoNED.

---

[10] We display results for our MATLAB fmincon implementation, which outperforms Kuosmanen and Kortelainen's (2012) results for a low noise-to-signal ratio, $\rho_{nts} = 1$, and has similar performance in Kuosmanen and Kortelainen's high noise-to-signal ratio, $\rho_{nts} = 2$. We did not compare performance when $\rho_{nts} = 3$, because Kuosmanen and Kortelainen do not estimate this scenario. See Kuosmanen and Kortelainen (2012) for a demonstration of StoNED's superior performance relative to standard implementations of SFA and DEA for all scenarios that included $\rho_{nts} > 0$.

*Example 2: Bivariate Cobb-Douglas frontier with homoscedastic inefficiency terms*

We fit the bivariate Cobb-Douglas frontier $Y_i = x_{1i}^{0.4} x_{2i}^{0.5} e^{-u_i} e^{v_i}$, where we consider a homoscedastic distribution for both noise and inefficiency, $u_i \sim \text{expo}(\mu_u = \sigma_u = 1/6)$ and $v_i \sim N(0, \sigma_v^2)$, where $\sigma_v = \rho_{nts} \sigma_u$[12], the same inefficiency assumptions as in Example 1. We also consider the same prior assumptions for $\theta$ as in Example 1. MBCR-I's estimators performance in terms of functional fit, MSE *f*, is better in all scenarios with a noise-to-signal ratio greater than 1. Despite StoNED's lower MSE *f* in the $\rho_{nts} = 1$ scenarios, MBCR-I's estimates give a more economically sound description of the frontier, because more of its hyperplanes are full-dimensional. In Example 2, non-full dimensional hyperplanes support between 14% and 23% of the observations for StoNED, whereas it is always less than 6% for MBCR-I. Finally, MBCR-I NS performs well when the number of observations is low and MBCR-I S well estimates inefficiency consistently.

---

[11] We define relatively constant as within a 5% difference across replicates for both the running mean and running standard deviation of *MSE f*.
[12] To make the CNLS problem feasible to solve, we reduced our optimality tolerance from 10$^{-10}$ to 10$^{-4}$ for the $n$ =500 scenarios on our GAMS with MINOS 5.5 implementation. Kuosmanen and Kortelainen (2012) did not perform $n$ =500 simulation scenarios in any of their multivariate examples.

Table 3 shows that predictions of the mean inefficiency are competitive for both StoNED and the MBCR-I estimators across the different scenarios varying the noise-to-signal ratio and number of observations, with the exception of the $\rho_{nts} = 3$ scenarios, where only MBCR-I S performs well. In Example 1 and the less noisy scenarios of Example 2, StoNED's functional estimates improve, as measured by $MSE\ f$, as the number of observations increases typical of any consistent estimator. However, in the high dimensionality and high noise scenarios for Example 2, StoNED's ability to fit the function, measured by $MSE\ f$, decreases as the number of observations increases. StoNED's erratic performance relates to the increase in local optima for the optimization problem associated with the first step of StoNED, CNLS, and occurs across all of the MATLAB and GAMS implementations. Even though we implement versions of CNLS which use global solvers,[13] the solvers cannot find solutions for data instances with more than 100 observations. MBCR-I, which uses an adaptive partitioning strategy rather than a full dataset optimization strategy does not suffer from these solution algorithm complexity issues.

Table 4 shows that the percentage of negatively skewed replicates also exhibits roughly consistent behavior throughout the different noise-to-signal ratio and the number of observations. However, for this more computationally challenging example, the percentage of negatively skewed replicates for StoNED is in general higher than in Example 1, and thus predicts negligible inefficiency levels more frequently. Further, both StoNED and MBCR-I need more replicates for their estimates to stabilize if compared with the simpler

---

[13] We attempted to use global nonlinear optimization algorithms such as MSNLP, BARON and ANTIGONE.

Example 1. However, we note that MBCR-I functional estimate converges in significantly less replicates than StoNED.

*Example 3: Trivariate Cobb-Douglas frontier with homoscedastic inefficiency terms*

We consider the Trivariate Cobb-Douglas frontier, $y_i = x_{1,i}^{0.4} x_{2,i}^{0.3} x_{3,i}^{0.2} e^{-u_i} e^{v_i}$. The distributional assumptions for the noise and inefficiency terms are the same as those of Examples 1 and 2, $u_i \sim \text{expo}(\mu_u = 1/6)$ and $v_i \sim N(0, \sigma_v^2)$. Prior assumptions on $\theta$ are the same as in previous examples. Table 5 shows that in this higher-dimensional setting, CNLS has a poor functional fit, MSE $f$, for scenarios with a large number of observations, $n = 500$, and for scenarios with a high noise-to-signal ratio, $\rho_{nts} = 3$.[14]

While the results for the noise-to-signal ratio equal to 1 scenarios are similar to Examples 1 and 2, with the MBCR-I estimators only being competitive in some of the scenarios in Example 3, the proportion of observations supported by non fully-dimensional hyperplanes fit by StoNED first-stage, CNLS, increases and ranges between 6% and 42%. For larger noise-to-signal ratios, $\rho_{nts} = 2,3$, the performance comparison is similar to Examples 1 and 2, with MBCR-I S performing the best in most of the scenarios. The variability in the functional fit, Standard Deviation of $MSE\ f$, and negative skewness results in Table 6 show behavior similar to Examples 1 and 2, with MBCR-I showing a lower inter-replicate variability in functional fit, $MSE\ f$.

*Example 4: Trivariate Cobb-Douglas frontier with heteroscedastic inefficiency terms*

---

[14] Again, the StoNED results from our GAMS with MINOS 5.5 implementation were similar or better than Kuosmanen and Kortelainen (2012). Kuosmanen and Kortelainen (2012) did not consider $n = 500$ scenarios. To make the CNLS problem feasible to solve, we reduced our optimality tolerance from $10^{-10}$ to $10^{-2}$ for the $n = 500$ scenarios.

We consider a heteroscedastic inefficiency DGP, where $u_i|\mathbf{x}_i \sim N^+(0, \sigma_{0u}(x_{1,i} + x_{2,i}))$, where $\sigma_u = 0.3$. The noise distribution is a homoscedastic Normal $v_i \sim N(0, \sigma_v^2)$, where $\sigma_v = \rho_{nts} \cdot \sigma_u \cdot \sqrt{(\pi-2)/\pi}$. The production frontier is $y_i = x_{1,i}^{0.4} x_{2,i}^{0.3} x_{3,i}^{0.2} e^{-u_i} e^{v_i}$, as in Example 3. The hyperparameter for our Exponential prior on inefficiency, $\theta$, is lower than in Examples 1, 2, and 3 due to the scale of the data, although still nearly uninformative with $v_0 = 1$ and $w_0$ drawn uniformly from the range (0, 0.1). Unlike Tables 1, 3, and 5, we include an additional set of results for StoNED from a different implementation.[15] We note that even if this alternative implementation has better functional fit, MSE $f$ has inconsistent behavior with regard to the percentage or negatively-skewed replicates.

Table 7 shows that the functional fit, MSE $f$, for the MBCR-I estimators is lower for all scenarios, with MBCR-I NS having significantly better performance when the number of observations is small and both MBCR-I estimators perform similarly for the larger $n$ scenarios. Reasons for MBCR-I's good functional fit are the updating of prior assumptions about the inefficiency term and incorporating hyperplane-specific noise variances into the posterior distribution of the observation's inefficiency term, $u_i$. Moreover, MBCR-I's mean inefficiency predictions are more accurate in 10 of the 12 scenarios. Conversely, StoNED's predictions for the mean inefficiency level are highly biased (even for $\rho_{nts} = 1$); however, this bias becomes smaller for our alternative implementation used in scenarios with 500 observations.

Comparing MBCR-I's results in Table 7 with MBCR-I's results for the correctly specified DGP in Table 5 shows that the heteroscedastic inefficiency specification only has a significantly detrimental impact on MSE $f$ for 4 of the 12 scenarios. Besides these, the MSE $f$ results for the heteroscedastic inefficiency example are at most 20% larger, and in fact smaller in the majority of scenarios. For both StoNED and MBCR-I, the percentage of observations supported by non fully-dimensional hyperplanes is similar to Example 3. We conclude that even for moderate sample sizes and large noise-to-signal ratios, MBCR-I is relatively robust to mis-specification of the inefficiency term. Finally, the DGP mis-specification impacts the variability of MBCR-I's functional fit across replicates less than StoNED's as measured by the Standard Deviation of MSE $f$ column in Table 8.

---

[15] The GAMS with MINOS 5.5 implementation is our main implementation for this example. The additional results are from our MATLAB fmincon implementation. Our main implementation was similar to the MSE $f$ results in Kuosmanen and Kortelainen (2012) for all $\rho_{nts} = 1$ scenarios. None of our implementations achieved Kuosmanen and Kortelainen (2012)'s $\rho_{nts} = 2, n = 300$ MSE $f$ results.

*3.2 Discussion of Simulation Results and recommendations to use MBCR-I*

As Tables 1–8 show, MBCR-I is best for scenarios with relatively noisy data and/or when the inefficiency distribution is unknown.[16] Relative to the benchmark method StoNED, MBCR-I is also competitive for lower noise-to-signal ratios in datasets where $n \geq 300$. Due to MBCR-I's one-stage nature, the residuals from this estimator are correctly skewed even for smaller datasets. Since the iteration-selection criterion is of no clear benefit for the sample sizes where MBCR-I is recommended, we choose the MBCR-I S estimator, because it eliminates the heuristic component of our algorithm. We also note that MBCR-I requires significantly higher computational times than CNLS for $n < 500$, partly because it gives a full posterior distribution of results rather than a point estimate. Nevertheless, due to its adaptive regression nature, MBCR-I only fits regression parameters for subsets of the dataset and so computational time increases slowly in $n$.[17] At $n = 500$, the computational time for both MBCR-I and Multiplicative-error StoNED is about 45 minutes.[18] MBCR-I is the only feasible existing axiomatic concavity constrained frontier estimation method to fit datasets with $1000 \leq n \leq 6000$ observations. Thus, we recommend MBCR-I S for most datasets with $300 \leq n \leq 6000$, or for smaller datasets when imposing the proper skewness on the residual distributions is needed. Oh et al. (2015) and Crispim Sarmento et al. (2015) analyze datasets where Multiplicative CNLS is not applicable due to dataset size, but both papers report successfully applying variants of MBCR-I.[19]

---

[16] An example of possible mis-specification of the inefficiency term appears in public sector applications where firms do not compete and efficient behavior does not result. Therefore, the distribution of inefficiency is unlikely to have a mode of zero and thus both an exponential or half-normal assumption regarding the inefficiency distributions is mis-specified.

We also recommend MBCR-I when inference on the frontier is needed. As a Bayesian method, MBCR-I produces credible intervals for MBCR-I S, our smoothed estimator, unlike CAP and StoNED, where it is computationally burdensome, even for moderate datasets, to obtain inference results by running the method repeatedly followed by bagging, smearing, or random partitioning (Hannah and Dunson, 2012). Finally, we recommend MBCR-I when obtaining a higher rate of fully dimensional hyperplanes is an important property of the frontier to be estimated.

## 4. Extensions

We consider several extensions to MBCR-I to make the method useful for a larger set of applications. We note that more general models for each extension are possible and are avenues for future research.

---

[17] MBCR-I's computational time increased from ~10 min for $n = 100$ to ~45 min for $n = 500$, whereas it increased from ~0.03min for $n = 100$ to ~45 min for $n = 500$ for Multiplicative CNLS.
[18] The CNLS speed-up algorithm proposed by Lee et al. (2013) on an additive error setting did not show time savings in our multiplicative setting.
[19] While MBCR-I integrates inefficiency into MBCR, it also has computational differences, as described in Section 2.5. Some of these are exploited to allow the inclusions of z-variables as discussed in Section 4.

*4.1 Flexible Time Trend*

MBCR-I is attractive for use with Panel data because it can fit shape constrained production frontiers for moderate datasets. To model technical progress over time, we consider a vector of dummy time effects to act as frontier-shifting factors (Baltagi and Griffin, 1988). We estimate the following model,

$$Y_{it} = f(\boldsymbol{X}_{it})e^{v_{it}}e^{-u_i}e^{\boldsymbol{\gamma}\boldsymbol{d}_{it}}, \qquad i = 1,..,n; \ t = 1,...T \qquad (11)$$

We let $\boldsymbol{\gamma} = (\gamma_2, ..., \gamma_T)$ and $\boldsymbol{d}_{it}$ is a row vector of dummy variables which has a 1 on the (*t*-1)*th* entry (and is a zero vector for observations on the first time period) and zeros on all other entries. Recalling MBCR-I's Gaussian mixture likelihood function, we know the hyperplane-specific noise variances $(\sigma_k^2)_{k=1}^K$ from the MBCR step of our algorithm (step 1). We let $\boldsymbol{D} = (\boldsymbol{d}_{11}, ..., \boldsymbol{d}_{nT})'$, collect the $(\sigma_{[i]}^2)_{i=1}^n$ terms on diagonal matrix $\Sigma_v$ and consider the Multivariate Normal prior $\boldsymbol{\gamma} \sim MVN(\mu_{\gamma 0}, \Sigma_{\gamma 0})$ to obtain the conjugate posterior shown in (12). Then, we add a step between steps 3 and 4 to draw $\boldsymbol{\gamma}^{(t)}$. Specifically,

$$\boldsymbol{\gamma}|... \sim MVN(\mu_{\gamma 1}, \Sigma_{\gamma 1}), \quad k = 1, ..., K, \text{where} \qquad (12)$$

$$\mu_{\gamma 1} = (\Sigma_{\gamma 0}^{-1} + \boldsymbol{D}'\Sigma_v^{-1}\boldsymbol{D})^{-1}(\Sigma_{\gamma 0}^{-1}\mu_{\gamma 0} + \boldsymbol{D}'\Sigma_v^{-1}\boldsymbol{D}\widehat{\boldsymbol{\gamma}})$$

$$\Sigma_{\gamma 1} = (\Sigma_{\gamma 0}^{-1} + \boldsymbol{D}'\Sigma_v^{-1}\boldsymbol{D})^{-1},$$

where $\widehat{\boldsymbol{\gamma}}$ is the OLS estimator of $r_i = ln(Y_i) - ln(\hat{f}_i) + u_i$ using $\boldsymbol{D}$ as a predictor. Since drawing from (12) is not computationally demanding, this estimator gives roughly the same computational performance as MBCR-I for $nT$ observations, thus opening the possibility to fit nonparametric multiplicative production frontiers to Panel datasets up to a few thousand observations. To adapt the MBCR-I algorithm to Panel data, we modify equations (9a) or (9b) to draw $(u_i)_{i=1}^n$ using information from the whole Panel rather than one observation. In the exponential inefficiency prior case, we consider the posterior hyperparameters analogous to O'Donnell and Griffiths (2006), who also consider a Gaussian mixture likelihood.

$$\sigma_{ui}^2 = \left[\sum_{t=1}^T \sigma_{[it]}^{-2}\right]^{-1}, \quad \mu_{ui} = \left\{\sum_{t=1}^T (\sigma_{[it]}^{-2})\left[\ln(\hat{f}_i) + \boldsymbol{\gamma}\boldsymbol{d}_{it} - \ln(Y_i)\right] - \theta\right\}\sum_{t=1}^T \sigma_{[it]}^{-2} \quad (13)$$

*4.2 Contextual Variables*

Incorporating contextual variables allows MBCR-I to estimate production functions that are multiplicatively affected by factors beyond a firm's control, or factors that are not inputs, but that the firm can control (see Johnson and Kuosmanen, 2011). We consider a parametric specification for the effect of contextual variables and fit the model

$$Y_i = f(\boldsymbol{X}_i)e^{v_i}e^{-u_i}e^{\boldsymbol{\delta}\boldsymbol{z}_i}, \qquad i = 1,..,n, \quad (14)$$

where $\boldsymbol{z}_{it} = (z_{1it},...,z_{Rit})'$ is the R-dimensional vector of z-variables for firm *i* at time *t* and $\boldsymbol{\delta} = (\delta_1,...,\delta_R)$ are the coefficients for each contextual variable. We note that (14) is the same as (11), with $\boldsymbol{\delta}$ and $\boldsymbol{z}_i$ playing the role of $\boldsymbol{\gamma}$ and $d_{it}$, respectively. Thus, the posterior simulation approach is the same as (12) with matrix $\boldsymbol{Z} = (\boldsymbol{z}'_{11},...,\boldsymbol{z}'_{nT})'$ in place of $\boldsymbol{D}$ and the prior parameters $(\mu_{\delta 0}, \Sigma_{\delta 0})$ in place of $(\mu_{\gamma 0}, \Sigma_{\gamma 0})$. We can also include both contextual variables and time effects with a common covariance matrix between their regression coefficients.

## 5. Empirical Application

Japan's recent plan for economic reforms, popularly termed "Abenomics", is lead by Prime Minister Shinzo Abe. The Prime Minister has recommended increased activity in the construction sector as an important economic driver, yet many politicians and pundits argue that Japan's construction industry is inefficient ("Japan and Abenomics", 2013). Our empirical application investigates the efficiency of operations in the Japanese concrete industry, a critical component of the country's construction industry.

We construct a dataset using the Census of Manufacturers collected by Japan's Ministry of Economy, Trade and Industry (METI) for concrete products. The data include all establishments in Japan with at least four workers.[20] We define Capital and Labor as the input variables and Value Added as the output variable. Capital and Value Added are measured in tens of thousands of Yen and Labor is measured in number of Employees. We report results for Cross Sectional datasets for 2007 and 2010 and a balanced Panel dataset for 2007-2010. See Appendix B for the Cross Sectional results for 2008 and 2009. We note even our smallest Cross Sectional dataset of 1,652 observations exceeds the capacity of existing methods to fit a semi-nonparametric shape constrained production function. Table 9 presents summary statistics.

---

[20] The equation used to calculate value added is different for firms with more than 30 employees than for firms with less than 30 employees. Excluding the firms with more than 30 employees did not result in significant changes to our frontier estimates, thus we present results for the full dataset.

We specify prior values for the parameters $K$, $(\sigma_k^2)_{k=1}^K$, $v_0$, $w_0$, $\theta$, $\boldsymbol{\gamma}$, $\boldsymbol{\delta}$. To determine the prior for $K$, we run multiplicative-error MBCR (without modeling inefficiency), and hypothesize that this prior value of $K$ is likely to be larger than the number of hyperplanes estimated by MBCR-I, because MBCR will capture more of the output variability with functional complexity, whereas MBCR-I can also use inefficiency. Unlike the $K = 1$ assumption, our MBCR-based prior on $K$ implies that we have the prior belief that the frontier has curvature and is more complex than a linear function. See Appendix C for a detailed discussion.[21] Additionally, we consider priors $v_0 = 1$, $w_0 = 0.1$ and $\theta = v_0 w_0$ for the inefficiency-related parameters. Finally, we consider a wide-support, nearly uninformative prior $IG(1, 0.01)$ for each $\sigma_k^2$. In the Panel model, we consider $\boldsymbol{\gamma}$ and $\boldsymbol{\delta}$ to have near-vague $MVN(0, M)$ prior distributions, where $M = diag_T(2000)$.

Due to our multiplicative error structure and the use of logarithms, we eliminate firms with negative Value Added (~1% of the initial observations). After an initial fit, 3% of the observations significantly deviate from our prediction. These observations correspond to the largest observations in terms of Capital, Value Added, or both, and thus we exclude these observations as outliers. Firm exclusion is consistent across all datasets.

*5.1 Posterior estimated frontier and interpretation*

Table 10 shows the fitting statistics for the two datasets. The number of hyperplanes needed by our Gaussian Mixture model is relatively small, which implies that firms operate in a few clusters with locally constant marginal productivity. We estimate a median inefficiency of 28-29% for the Cross Sectional and 45% for the Panel datasets, respectively. The percentage of output variability explained by the joint production function and inefficiency model is between 66-76% for the Cross Sectional and above 80% for the Panel datasets, respectively. The number of hyperplanes fitted for the Panel dataset is larger, due to both MBCR-I's ability to produce finer estimates of the production function when more data is available, and to the lower noise level of the firms operating throughout the study period. We report the percentage of output variation explained by our model and the part that remains unexplained we report as noise in the columns labeled % Model and % Noise, respectively.

---

[21] This assumption did not have a significant impact on our parameter estimates compared to the $K = 1$ assumption, but helped MBCR-I to be more computationally efficient and resulted in the insights summarized in *Appendix C*.

Tables 11a, 11b, and 11c show the economic quantities of interest for MBCR-I S, i.e., the marginal productivities of Capital and Labor, Capital to Labor Elasticity of Substitution and Technical Efficiency of the fitted production frontier models according to their minimum, maximum, and quartile-specific values across observations. Table 11c, however, shows Technical Efficiencies on a firm-specific basis, rather than an observation-specific basis. While we observe non-constant elasticities of substitution across the samples, Tables 11a and 11b show that a single component of their respective Gaussian mixtures support at least half of the observations, indicating that a majority group faces the same marginal productivity and Elasticity of Substitution. Regarding Technical Efficiency, in 2010 413 out of 1,652 establishments (25%) operate below 53% efficiency. This level of inefficiency is consistent across all years and the Panel dataset analysis, indicating a large potential for improvement for this subset of establishments. The Cross Sectional results for 2007 and 2010 (and the 2008 and 2009 results in Appendix B), are similar in most of the estimated frontier and efficiency characteristics, indicating firm's consistency performance

over time. Table 11c shows that the Panel data model gives a smoother, more detailed estimate of the frontier.

The lower Technical Efficiency levels for the Panel model are conservative estimates, because as Table 12 shows, the frontier contracts in each time period after 2007are likely due to the global financial crisis. Given that 2007 is the base year for the Panel analysis and that our assumption about uniform yearly frontier shifts across firms is an approximation, we suggest that firm-specific inefficiency terms absorb some of the frontier shrinkage effect. Finally, we note that the frontier multiplier for year *t* is given by $e^{\gamma_t}$, as defined in Section 4.1.Table 12 shows Maximum a Posteriori (MAP)[22] and 90% credible interval values for the year-specific frontier shifts for the Panel model.

Table 13 shows the Most Productive Scale size results, conditional on Capital/Labor ratio distributions at the 10, 25, 50, 75, and 90 percentiles (see Appendix D for a full discussion). Firm-specific technical efficiencies are lower at all quantiles for Cross Sectional versus the Panel data results. The Capital/Labor ratio distribution is similar for the Cross Sectional and Panel datasets. The MPSS results for the Panel dataset are similar to those of 2007, the reference year for the Panel, which is another indication of consistency across the estimated frontiers. The Cross Sectional results show higher MPSS for 2010, suggesting technical progress over the timespan. Table 12 also confirms that the global financial crisis of 2007 has a significant effect on the future output of establishments.

---

[22] The MAP value is the highest density point of the simulated posterior distribution for the parameter of interest.

## 6. Conclusions

This paper described MBCR-I, a one-stage Bayesian semi non-parametric method to fit concave and monotonic stochastic frontiers using Reversible Jump Markov Chain Monte Carlo techniques. Computational modifications allowed for both the presence of inefficiency and the multiplicative residuals structure of the standard SFA model. MBCR-I's performance in the Monte Carlo simulation study showed increased accuracy in large noise-to-signal ratio scenarios versus StoNED and for large sample sizes at any noise-to-signal ratio. MBCR-I handled datasets of a few thousand observations efficiently, which suggested its use for long panels of moderate samples and very large cross sections. MBCR-I was increasingly robust to mis-specification of the inefficiency model due to its ability to learn from data and to consider locally shrunk individual inefficiency posteriors.

MBCR-I was empirically tested by using data on Japan's concrete firms operating between 2007 and 2010. Computation of input productivities, elasticities of substitution, inefficiency distributions, frontier-shifting effects and most productive scale sizes were demonstrated. MBCR-I's Bayesian framework made straightforward inference possible for the frontier-shifting effects.

There was limited evidence for the criticism that important parts of the construction industry offered significant room for efficiency improvements. Between 2007–2010, efficiency levels were stable at relatively high levels. Japan was significantly affected by the global financial crisis (Fukao and Yuan, 2009) as shown by the value-added output steadily declined over the 2007-2010 time period.

We believe that the work described in this paper is the first model that allows a shaped constrained production frontier to be estimated nonparametrically and relaxes the homoscedastic assumption on the inefficiency term for the analysis of cross sectional data.

Our experiments revealed the tradeoffs between modeling the frontier and the inefficiency term, i.e., more flexible models of the inefficiency term lead to coarser estimates of the production frontier and vice versa.

Future research should consider a more extensive exploration of the tradeoffs. A second research path should consider developing contextual variable models that do not rely on parametric assumptions. While we obtained credible intervals for all estimated parameters, given our non-linear least squares step to draw the hyperplane-specific regression coefficients, our credible intervals on both the production frontier and contextual variables were conditional on the point estimates obtained in that step. Although this assumption is not unreasonable, because it considers the best hyperplane fits conditional on all other parameters, future research, should also extend MBCR-I to draw the regression coefficients with a more efficient Bayesian algorithm in order to obtain full posterior distributions (and credible intervals) for the estimated parameters. Compared with StoNED, MBCR-I fitted a very small number of hyperplanes on highly clustered datasets, especially for low-dimensional input vectors. Thus, future research on alternative strategies for proposing knots and directions when adding hyperplanes should lead to more detailed frontier estimations.

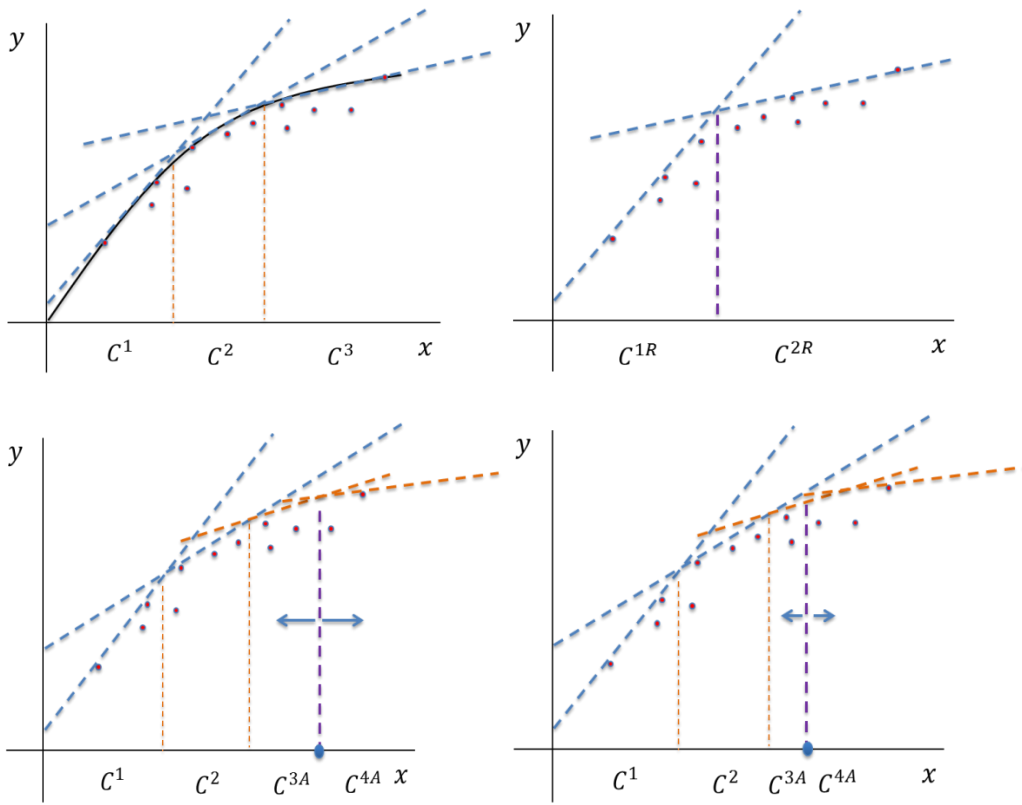

**Figure 1. Three basis regions defined by the current number of hyperplanes $K^{(t)} = 3$ (top left). Proposed basis regions for removal of the second hyperplane (top right). Proposal basis regions for hyperplane addition and split of third basis region (bottom left). Proposal basis regions for hyperplane addition and split of third basis region using a different splitting knot and the same splitting direction (bottom right).**

**Table 1. Results for Example 1: Univariate Cobb-Douglas frontier with homoscedastic inefficiency terms**

| $\rho_{nts}$ | $n$ | MSE $f$ | | | $\widehat{E(\bar{u})} - E(\bar{u})$ | | | % Non-Full Dimensional | |
|---|---|---|---|---|---|---|---|---|---|
| | | StoNED | MBCR-INS | MBCR-IS | StoNED | MBCR-INS | MBCR-IS | StoNED | MBCR-INS |
| 1 | 100 | **0.0035** | 0.0046 | 0.0036 | **0.01** | **0.01** | **0.01** | 4% | **0%** |
| | 200 | **0.0010** | 0.0020 | 0.0024 | **0.01** | **0.01** | **0.01** | 2% | **0%** |
| | 300 | 0.0009 | 0.0014 | **0.0007** | **0.01** | **0.01** | **0.01** | 3% | **0%** |
| | 500 | **0.0005** | 0.0008 | 0.0006 | **0.01** | **0.01** | **0.01** | 1% | **0%** |
| 2 | 100 | 0.0054 | 0.0056 | **0.0035** | 0.02 | -0.02 | -0.02 | 6% | **0%** |
| | 200 | 0.0068 | 0.0052 | **0.0035** | 0.05 | -0.03 | -0.03 | 2% | **0%** |
| | 300 | 0.0062 | 0.0055 | **0.0028** | 0.05 | **0.01** | -0.05 | 3% | **0%** |
| | 500 | 0.0068 | 0.0049 | **0.0023** | 0.07 | 0.05 | **-0.02** | 3% | **0%** |
| 3 | 100 | 0.0334 | **0.0097** | 0.0153 | 0.09 | **0.00** | -0.01 | 8% | **0%** |
| | 200 | 0.0464 | **0.0054** | 0.0061 | 0.18 | **0.02** | **0.02** | 7% | **0%** |
| | 300 | 0.0420 | 0.0059 | **0.0046** | 0.18 | **-0.03** | -0.04 | 8% | **0%** |
| | 500 | 0.0413 | 0.0050 | **0.0028** | 0.16 | **-0.06** | -0.07 | 2% | **0%** |

**Table 2. Estimator Robustness Analysis for Example 1: Univariate Cobb-Douglas frontier with homoscedastic inefficiency terms**

| $\rho_{nts}$ | $n$ | Replicates for MSE $f$ convergence | | Standard Deviation of MSE $f$ | | MSE $f$ coefficient of variation | | % Negative Skew |
|---|---|---|---|---|---|---|---|---|
| | | StoNED | MBCR-I | StoNED | MBCR-I | StoNED | MBCR-I | StoNED |
| 1 | 100 | 10 | 10 | 0.0046 | **0.0019** | 132% | **42%** | 10% |
| | 200 | 10 | 10 | **0.0011** | 0.0016 | 108% | **64%** | 0% |
| | 300 | 10 | 20 | **0.0009** | 0.0018 | **101%** | 128% | 0% |
| | 500 | 10 | 10 | **0.0005** | **0.0005** | 85% | **63%** | 0% |
| 2 | 100 | 10 | 10 | 0.0056 | **0.0036** | 104% | **64%** | 10% |
| | 200 | 10 | 10 | 0.0058 | **0.0018** | 85% | **34%** | 0% |
| | 300 | 10 | 10 | **0.0022** | 0.0038 | **35%** | 69% | 0% |
| | 500 | 10 | 10 | 0.0048 | **0.0047** | **71%** | 95% | 0% |
| 3 | 100 | 10 | 20 | 0.0246 | **0.0077** | 74% | 75% | 20% |
| | 200 | 10 | 10 | 0.0278 | **0.0024** | 60% | **45%** | 0% |
| | 300 | 10 | 10 | 0.0167 | **0.0019** | 40% | **37%** | 0% |
| | 500 | 10 | 10 | 0.0142 | **0.0033** | **34%** | 57% | 0% |

**Table 3. Results for Example 2: Bivariate Cobb-Douglas frontier with homoscedastic inefficiency terms**

| $\rho_{nts}$ | $n$ | MSE $f$ | | | $\widehat{E(\bar{u})} - E(\bar{u})$ | | | % Non Full-Dimensional | |
|---|---|---|---|---|---|---|---|---|---|
| | | StoNED | MBCR-INS | MBCR-IS | StoNED | MBCR-INS | MBCR-IS | StoNED | MBCR-INS |
| 1 | 100 | **0.0022** | 0.0032 | 0.0038 | **-0.02** | -0.07 | 0.03 | 16% | **2%** |
| | 200 | **0.0015** | **0.0015** | 0.0018 | -0.02 | **-0.01** | 0.04 | 15% | **0%** |
| | 300 | **0.0014** | 0.0020 | 0.0016 | -0.02 | **-0.01** | 0.02 | 17% | **0%** |
| | 500 | **0.0004** | 0.0010 | 0.0012 | **-0.01** | 0.01 | 0.02 | 10% | **6%** |
| 2 | 100 | 0.0058 | 0.0043 | **0.0040** | **0.02** | -0.09 | **0.01** | 18% | **1%** |
| | 200 | 0.0051 | **0.0039** | 0.0050 | **0.01** | -0.04 | 0.06 | 19% | **0%** |
| | 300 | 0.0045 | 0.0040 | **0.0036** | **0.01** | -0.08 | 0.03 | 20% | **0%** |
| | 500 | 0.0047 | 0.0029 | **0.0014** | 0.07 | -0.08 | **0.00** | 16% | **0%** |
| 3 | 100 | 0.0237 | **0.0085** | 0.0175 | 0.10 | **-0.06** | 0.12 | 23% | **0%** |
| | 200 | 0.0259 | **0.0045** | 0.0057 | 0.14 | -0.11 | **0.00** | 23% | **5%** |
| | 300 | 0.0306 | 0.0050 | **0.0046** | 0.17 | -0.11 | **-0.02** | 22% | **2%** |
| | 500 | 0.0270 | 0.0047 | **0.0032** | 0.13 | -0.12 | **-0.06** | 14% | **3%** |

**Table 4. Estimator Robustness Analysis for Example 2: Bivariate Cobb-Douglas frontier with homoscedastic inefficiency terms**

|  |  | Replicates for MSE $f$ convergence |  | Standard Deviation of MSE $f$ |  | MSE $f$ coefficient of variation |  | % Negative Skew |
|---|---|---|---|---|---|---|---|---|
| $\rho_{nts}$ | $n$ | StoNED | MBCR-I | StoNED | MBCR-I | StoNED | MBCR-I | StoNED |
| 1 | 100 | 50 | 10 | 0.0024 | **0.0013** | 113% | **40%** | 10% |
|   | 200 | 50 | 10 | 0.0017 | **0.0007** | 120% | **48%** | 10% |
|   | 300 | 50 | 40 | 0.0020 | **0.0013** | 145% | **65%** | 10% |
|   | 500 | 10 | 20 | **0.0002** | 0.0007 | **56%** | 69% | 0% |
| 2 | 100 | 20 | 10 | 0.0032 | **0.0017** | 56% | **40%** | 15% |
|   | 200 | 20 | 10 | 0.0038 | **0.0017** | 75% | **43%** | 20% |
|   | 300 | 20 | 20 | 0.0024 | **0.0015** | 54% | **39%** | 20% |
|   | 500 | 10 | 20 | 0.0023 | **0.0017** | **49%** | 58% | 0% |
| 3 | 100 | 20 | 10 | 0.0164 | **0.0045** | 69% | **53%** | 15% |
|   | 200 | 20 | 10 | 0.0130 | **0.0012** | 50% | **27%** | 5% |
|   | 300 | 20 | 20 | 0.0120 | **0.0024** | **39%** | 48% | 0% |
|   | 500 | 10 | 10 | 0.0141 | **0.0013** | 52% | **34%** | 10% |

**Table 5. Results for Example 3: Trivariate Cobb-Douglas frontier with homoscedastic inefficiency terms**

| $\rho_{nts}$ | $n$ | MSE $f$ | | | $\widehat{E(\tilde{u})} - E(\bar{u})$ | | | % Non Full-Dimensional | |
| --- | --- | --- | --- | --- | --- | --- | --- | --- | --- |
| | | StoNED | MBCR-INS | MBCR-IS | StoNED | MBCR-INS | MBCR-IS | StoNED | MBCR-INS |
| 1 | 100 | **0.0015** | 0.0055 | 0.0056 | **-0.01** | **-0.01** | 0.07 | 33% | **5%** |
| | 200 | **0.0014** | 0.0030 | 0.0033 | -0.02 | **0.01** | 0.06 | 25% | **12%** |
| | 300 | **0.0020** | 0.0026 | 0.0028 | **0.02** | **0.02** | 0.05 | 23% | **1%** |
| | 500 | 0.0070 | 0.0019 | **0.0013** | -0.08 | **0.01** | 0.03 | 6% | **0%** |
| 2 | 100 | **0.0053** | 0.0078 | 0.0072 | **0.00** | -0.07 | 0.04 | 38% | **8%** |
| | 200 | **0.0054** | 0.0060 | 0.0070 | **0.03** | -0.07 | 0.09 | 38% | **10%** |
| | 300 | 0.0075 | 0.0047 | **0.0045** | 0.07 | **0.04** | **0.04** | 34% | **4%** |
| | 500 | 0.0147 | 0.0039 | **0.0031** | 0.14 | -0.07 | **0.01** | **6%** | **6%** |
| 3 | 100 | 0.0286 | **0.0076** | 0.0088 | 0.12 | -0.10 | **0.01** | 42% | **2%** |
| | 200 | 0.0243 | 0.0077 | **0.0063** | 0.12 | -0.10 | **0.01** | 38% | **1%** |
| | 300 | 0.0218 | 0.0058 | **0.0042** | 0.13 | -0.10 | **-0.02** | 32% | **3%** |
| | 500 | 0.0257 | 0.0066 | **0.0034** | 0.11 | -0.12 | **-0.04** | 14% | **0%** |

**Table 6. Estimator Robustness Analysis for Example 3: Trivariate Cobb-Douglas frontier with homoscedastic inefficiency terms**

| $\rho_{nts}$ | $n$ | Replicates for MSE $f$ convergence | | Standard Deviation of MSE $f$ | | MSE $f$ coefficient of variation | | % Negative Skew |
|---|---|---|---|---|---|---|---|---|
| | | StoNED | MBCR-I | StoNED | MBCR-I | StoNED | MBCR-I | StoNED |
| 1 | 100 | 20 | 10 | **0.0006** | 0.0020 | 39% | **37%** | 0% |
| | 200 | 50 | 10 | **0.0010** | 0.0012 | 71% | **41%** | 4% |
| | 300 | 50 | 20 | 0.0033 | **0.0011** | 163% | **43%** | 2% |
| | 500 | 10 | 10 | 0.0054 | **0.0008** | 76% | **40%** | 0% |
| 2 | 100 | 20 | 10 | **0.0025** | 0.0027 | 48% | **34%** | 15% |
| | 200 | 20 | 10 | 0.0025 | **0.0017** | 46% | **29%** | 10% |
| | 300 | 20 | 20 | 0.0065 | **0.0021** | 87% | **47%** | 0% |
| | 500 | 10 | 20 | 0.0057 | **0.0015** | **38%** | 38% | 0% |
| 3 | 100 | 50 | 20 | 0.0204 | **0.0037** | 71% | **48%** | 10% |
| | 200 | 20 | 10 | 0.0129 | **0.0031** | 53% | **41%** | 10% |
| | 300 | 20 | 10 | 0.0130 | **0.0022** | 60% | **39%** | 10% |
| | 500 | 10 | 10 | 0.0197 | **0.0026** | 77% | **40%** | 10% |

**Table 7. Results for Example 4: Trivariate Cobb-Douglas frontier with heteroscedastic inefficiency terms**

| | | MSE $f$ | | | $\widehat{E(\bar{u})} - E(\bar{u})$ | | | % Non Full-Dimensional | |
|---|---|---|---|---|---|---|---|---|---|
| $\rho_{nts}$ | $n$ | StoNED | MBCR-I NS | MBCR-I S | StoNED | MBCR-I NS | MBCR-I S | StoNED | MBCR-I NS |
| 1 | 100 | 0.0058 | **0.0046** | 0.0071 | 0.10 | **0.02** | 0.11 | 25% | **6%** |
| | 200 | 0.0047 | **0.0025** | 0.0033 | 0.10 | **0.02** | 0.07 | 24% | **4%** |
| | 300 | 0.0047 | **0.0033** | 0.0045 | 0.10 | **0.06** | 0.10 | 23% | **1%** |
| | 500 | 0.0499, 0.0028 | **0.0015** | 0.0019 | 0.33, **0.02** | 0.04 | 0.04 | 27% | **1%** |
| 2 | 100 | 0.0507 | **0.0127** | 0.0194 | 0.28 | **0.12** | 0.22 | 45% | **0%** |
| | 200 | 0.0407, 0.0158 | **0.0064** | 0.0126 | 0.27, 0.08 | **0.07** | 0.16 | 35% | **5%** |
| | 300 | 0.0590, 0.0191 | **0.0042** | 0.0052 | 0.33, 0.12 | **0.00** | 0.06 | 31% | **3%** |
| | 500 | 0.1345, 0.0084 | **0.0020** | 0.0023 | 0.47, 0.05 | **0.03** | 0.03 | 12% | **0%** |
| 3 | 100 | 0.1492 | **0.0107** | 0.0205 | 0.38, 0.28 | **0.04** | 0.18 | 48% | **0%** |
| | 200 | 0.1955, 0.1286 | **0.0061** | 0.0068 | 0.50, 0.33 | **0.03** | 0.06 | 47% | **4%** |
| | 300 | 0.2446, 0.1046 | **0.0050** | 0.0050 | 0.58, 0.29 | **0.00** | 0.04 | 30% | **5%** |
| | 500 | 0.2610, 0.0061 | **0.0034** | 0.0038 | 0.63, **-0.02** | 0.04 | 0.04 | 20% | **5%** |

**Table 8. Estimator Robustness Analysis for Example 3: Trivariate Cobb-Douglas frontier with heteroscedastic inefficiency terms**

| $\rho_{nts}$ | $n$ | Replicates for MSE $f$ convergence | | Standard Deviation of MSE $f$ | | MSE $f$ coefficient of variation | | % Neg. Skew |
|---|---|---|---|---|---|---|---|---|
| | | StoNED | MBCR-I | StoNED | MBCR-I | StoNED | MBCR-I | StoNED |
| 1 | 100 | 20 | 10 | 0.0041 | **0.0027** | 71% | **55%** | 15% |
| | 200 | 20 | 10 | 0.0023 | **0.0011** | 48% | **44%** | 20% |
| | 300 | 20 | 10 | 0.0028 | **0.0017** | 59% | **51%** | 10% |
| | 500 | 10 | 10 | 0.0314, 0.0019 | **0.0008** | 63%, 66% | **58%** | 0%, 20% |
| 2 | 100 | 20 | 10 | 0.0405 | **0.0079** | 80% | **62%** | 5% |
| | 200 | 20 | 10 | 0.0233, 0.0216 | **0.0051** | **57%**, | 81% | 10%, 70% |
| | 300 | 30 | 10 | 0.0399, 0.0173 | **0.0032** | **68%**, | 75% | 5%, 50% |
| | 500 | 10 | 20 | 0.0826, 0.0122 | **0.0005** | 61%, 146% | **26%** | 10%, 70% |
| 3 | 100 | 50 | 20 | 0.1277, 0.1428 | **0.0067** | 86%, 108% | **63%** | 20% |
| | 200 | 50 | 10 | 0.0954, 0.1088 | **0.0018** | 49%, 85% | **30%** | 5%, 30% |
| | 300 | 40 | 10 | 0.1298, 0.1010 | **0.0026** | 53%, 142% | **51%** | 3%, 40% |
| | 500 | 10 | 10 | 0.0641, 0.0028 | **0.0011** | **25%**, 46% | 33% | 0%, 100% |

**Table 9. Descriptive Statistics for Concrete Products Dataset**

| Year | | 2007 | 2008 | 2009 | 2010 |
|---|---|---|---|---|---|
| Number of observations – Cross Sectional | | 1,929 | 1,715 | 1,714 | 1,652 |
| Capital | Mean | 16,419 | 15,809 | 15,662 | 14,215 |
| | Median | 2,000 | 2,000 | 2,000 | 2,000 |
| | Standard Deviation | 76,401 | 74,300 | 69,332 | 58,486 |
| Labor | Mean | 16.68 | 17.51 | 16.60 | 16.32 |
| | Median | 12 | 13 | 12 | 12 |
| | Standard Deviation | 15.80 | 15.83 | 15.39 | 15.29 |
| Value Added | Mean | 19,004 | 18,481 | 18,393 | 16,520 |
| | Median | 11,515 | 11,391 | 11,092 | 10,587 |
| | Standard Deviation | 26,221 | 21,619 | 20,842 | 17,407 |

| Year | | 2007 | 2008 | 2009 | 2010 |
|---|---|---|---|---|---|
| Number of observations – Panel | | 1,382 | 1,382 | 1,382 | 1,382 |
| Capital | Mean | 16,313 | 16,475 | 15,901 | 15,536 |
| | Median | 2,000 | 2,000 | 2,000 | 2,000 |
| | Standard Deviation | 77,269 | 77,426 | 70,606 | 63,081 |
| Labor | Mean | 18.44 | 18.25 | 17.72 | 17.09 |
| | Median | 14 | 14 | 14 | 13 |
| | Standard Deviation | 16.38 | 16.20 | 15.91 | 15.75 |
| Value Added | Mean | 20,084 | 19,434 | 19,377 | 17,648 |
| | Median | 13,502 | 12,385 | 12,548 | 11,795 |
| | Standard Deviation | 22,475 | 20,264 | 20,665 | 17,839 |

**Table 10. MBCR-I S Cross Sectional and Panel production frontier fitting statistics**

|       | # Hyperplanes fitted | Median Inefficiency | % Model | % Noise |
|-------|---------------------|---------------------|---------|---------|
| 2007  | 2.11                | 29%                 | 66%     | 34%     |
| 2008  | 2.03                | 28%                 | 75%     | 25%     |
| 2009  | 2.60                | 29%                 | 70%     | 30%     |
| 2010  | 2.36                | 28%                 | 76%     | 24%     |
| Panel | 3.63                | 45%                 | 81%     | 19%     |

**Table 11a. MBCR-I S Cross Sectional production frontier characterization for 2007**

|  | Marginal Product | | Elasticity of Substitution (x$10^{-4}$) | Technical Efficiency |
| --- | --- | --- | --- | --- |
|  | Capital | Labor | Capital/Labor | Firm-specific % |
| Min | 0.0446 | 1,013 | 0.4278 | 0.060% |
| 25th percentile | 1.0534 | 1,222 | 8.6449 | 50.89% |
| Median | 1.0534 | 1,222 | 8.6449 | 70.94% |
| 75th percentile | 1.0534 | 1,222 | 8.6449 | 86.61% |
| Max | 3.7182 | 1,218 | 43.189 | 99.97% |

**Table 11b. MBCR-I S Cross Sectional production frontier characterization for 2010**

|  | Marginal Product | | Elasticity of Substitution (x$10^{-4}$) | Technical Efficiency |
| --- | --- | --- | --- | --- |
|  | Capital | Labor | Capital/Labor | Firm-specific % |
| Min | 0.0289 | 803.1 | 0.3342 | 3.46% |
| 25th percentile | 0.9880 | 1074 | 9.2108 | 52.98% |
| Median | 0.9880 | 1074 | 9.2108 | 72.80% |
| 75th percentile | 0.9880 | 1074 | 9.2108 | 87.66% |
| Max | 6.7196 | 1074 | 98.7883 | 97.97% |

**Table 11c. MBCR-I Panel production frontier characterization**

|  | Marginal Product | | Elasticity of Substitution | Technical Efficiency |
| --- | --- | --- | --- | --- |
|  | Capital | Labor | Capital/Labor (x$10^{-4}$) | Firm-specific % |
| Min | 0.0299 | 988 | 0.2850 | 2.78% |
| 25th percentile | 0.8758 | 1173 | 7.2168 | 36.53% |
| Median | 1.8886 | 1318 | 14.0175 | 55.40% |
| 75th percentile | 3.0354 | 1417 | 21.4405 | 77.80% |
| Max | 3.7157 | 1971 | 26.3485 | 99.95% |

**Table 12. Time dummy variable coefficients**

|  | 2008 | 2009 | 2010 |
|---|---|---|---|
| MAP | -0.0595 | -0.0628 | -0.1063 |
| Frontier Multiplier | 0.9422 | 0.9391 | 0.8992 |
| 90% Credible Interval | (0.9186, 0.9665) | (0.9154, 0.9649) | (0.8762, 0.9248) |

**Table 13. Most productive scale size for selected Capital/Labor ratios for Cross Sectional and Panel production frontiers.**

| Model | Capital/Labor ratio percentile | 10% | 25% | 50% | 75% | 90% |
|---|---|---|---|---|---|---|
| Cross Sectional 2007 | Capital/Labor | 37 | 75 | 166.7 | 400 | 1,500 |
| | MPSS Capital | 2,837 | 4,744 | 7,489 | 10,675 | 14,791 |
| | MPSS Labor | 76 | 63 | 44 | 27 | 10 |
| Cross Sectional 2010 | Capital/Labor | 37.5 | 80.0 | 166.7 | 416.7 | 1451.2 |
| | MPSS Capital | 3,662 | 6,114 | 9,565 | 13,965 | 19,142 |
| | MPSS Labor | 94 | 79 | 59 | 36 | 13 |
| Panel 2007–2010 | Capital/Labor | 38.9 | 76.9 | 162.2 | 389.9 | 1456.0 |
| | MPSS Capital | 2,210 | 3,835 | 6,241 | 9,546 | 14,881 |
| | MPSS Labor | 57 | 50 | 38 | 24 | 10 |

## Appendix A. Fully Bayesian MBCR-I

As mentioned in Section 2.3, we solve mathematical program (7) instead of simulating $(\alpha_k, \beta_k)_{k=1}^K$ from its posterior distribution due to the difficulty of obtaining good proposal distributions. Computational feasibility of the original, additive error structure, MCBR algorithm depends on the availability of good proposal distributions for $(\alpha_k, \beta_k)_{k=1}^K$ (Hannah and Dunson, 2011). While H-D take advantage of conjugacy to compute such proposal distributions, we are not able to do so due to the logarithm operator present in our likelihood function, $\prod_{i=1}^n N\left(\ln(Y_i) - \ln(\alpha_{[i]} + \beta_{[i]}^T X_i) + u_i, \sigma_{[i]}^2\right)$. Here, we propose a naïve way to obtain full conditional posterior distributions for $(\alpha_k, \beta_k)_{k=1}^K$. The algorithm to draw $(\alpha_k, \beta_k)$ for each basis region is as follows:

1. Solve mathematical program (7) to obtain candidate distribution mean estimates $(\alpha_{k1}, \beta_{k1})$ for the given basis region.

2. Draw the candidate values $(\alpha_{kCand}, \beta_{kCand})$ from a $(d+1)$-dimensional Multivariate Gaussian distribution $N_{(d+1)}((\alpha_{k1}, \beta_{k1}), \Sigma_{\alpha,\beta})$, where $\Sigma_{\alpha,\beta} = \eta \begin{bmatrix} \sigma_{\alpha_k}^2 & \cdots & 0 \\ \vdots & \ddots & \vdots \\ 0 & \cdots & \sigma_{\beta_{dk}}^2 \end{bmatrix}$, where $\eta > 1$ is a tunable parameter, and $\sigma_{\alpha_k}, \ldots, \sigma_{\beta_{dk}}^2$ are the standard errors of $\alpha_{k1}$ and each component of $\beta_{k1}$ obtained after solving (7).

3. Draw $\sigma_{kCand}^2$ from (8), assuming $(\alpha_{kCand}, \beta_{kCand})$.

4. Compute the Metropolis-Hastings acceptance probability $a_{Prob}$ for $(\alpha_{kCand}, \beta_{kCand}, \sigma_{kCand}^2)$ using (A1)-(A4):

(A1) $a_{Prob} = \dfrac{p(\alpha_{kCand}, \beta_{kCand}, \sigma^2_{kCand}) L_k(\alpha_{kCand}, \beta_{kCand}, \sigma^2_{kCand})}{q(\alpha_{kCand}, \beta_{kCand}, \sigma^2_{kCand} | \alpha_{kCurr}, \beta_{kCurr}, \sigma^2_{kCurr})} \bigg/ \dfrac{p(\alpha_{kCurr}, \beta_{kCurr}, \sigma^2_{kCurr}) L_k(\alpha_{kCurr}, \beta_{kCurr}, \sigma^2_{kCurr})}{q(\alpha_{kCurr}, \beta_{kCurr}, \sigma^2_{kCurr} | \alpha_{kCand}, \beta_{kCand}, \sigma^2_{kCand})}$

(A2) $p(\alpha, \beta, \sigma^2) \sim N_{(d+1)} IG(\mathbf{0}, MI, \tilde{a}, \tilde{b})$, where $M$ is a large number and $I$ is the identity matrix

(A3) $L_k(\alpha, \beta, \sigma^2) = \prod_{i=1}^{n_k} N(\ln(Y_i) - \ln(\alpha + \beta^T X_i) + u_i, \sigma^2)$

(A4) $q(\alpha_2, \beta_2, \sigma^2_2 | \alpha_1, \beta_1, \sigma^2_1) \sim N_{(d+1)} IG(\mu_1, \Sigma_1, a_1, b_1)$, where $(\mu_1, \Sigma_1, a_1, b_1)$ are the hyperparameters associated with $(\alpha_1, \beta_1, \sigma^2_1)$.

5. If $a_{Prob} > Unif(0,1)$, accept draw $(\alpha_{kCand}, \beta_{kCand}, \sigma^2_{kCand})$; otherwise, go back to step 2.

6. After accepting a predefined number $n_{bi}$ of burn-in draws, accept the $(n_{bi} + 1)th$ accepted draw as a valid draw from the posterior distribution of $(\alpha_k, \beta_k, \sigma^2_k)$.

**Appendix B. Frontier characteristics for 2008-2009 Cross Sectional datasets**

Tables B1 and B2 show that the Marginal Product and Technical Efficiency quantiles are similar to those of 2010 for almost all cases, which is also true for the Elasticity of Substitution between Capital and Labor, except for its maximum values. In the case of MPSS shown in Table B3, the only significant departure from the results in 2007 and 2010 are those corresponding to the lower Capital/Labor ratios in the 2008 dataset. Observing that 2008 has the simplest frontier shape (2.03 hyperplanes) suggests that the data in 2008 may not be as well-behaved or as clean as in the other years. Note that the MPSS for Capital does not follow the monotonic trend observed for all other years and for the Panel dataset.

**Table B1. MBCR-I S Cross Sectional production frontier characterization for 2008**

|  | Marginal Product | | Elasticity of Substitution ($\times 10^{-4}$) | Technical Efficiency |
|---|---|---|---|---|
|  | Capital | Labor | Capital/Labor | Firm-specific % |
| Min | 0.0093 | 1,119 | 0.0767 | 1.26% |
| 25th percentile | 1.1209 | 1,202 | 9.3052 | 51.90% |
| Median | 1.1209 | 1,206 | 9.3052 | 71.77% |
| 75th percentile | 1.1209 | 1,206 | 9.3052 | 87.07% |
| Max | 1.4449 | 1,262 | 13.142 | 99.97% |

**Table B2. MBCR-I S Cross Sectional production frontier characterization for 2009**

|  | Marginal Product | | Elasticity of Substitution (x10$^{-4}$) | Technical Efficiency |
|---|---|---|---|---|
|  | Capital | Labor | Capital/Labor | Firm-specific % |
| Min | 0.0662 | 910 | 0.5608 | 0.97% |
| 25th percentile | 1.0010 | 1,191 | 8.4305 | 50.82% |
| Median | 1.0010 | 1,191 | 8.4305 | 71.04% |
| 75th percentile | 1.0012 | 1,191 | 8.4328 | 86.72% |
| Max | 10.0217 | 1,220 | 134.53 | 99.97% |

**Table B3. Most productive scale size for selected Capital/Labor ratios for Cross Sectional frontiers 2008-2009.**

| Model | Capital/Labor ratio percentile | 10% | 25% | 50% | 75% | 90% |
|---|---|---|---|---|---|---|
| Cross Sectional 2008 | Capital/Labor | 38 | 75 | 166.7 | 400 | 1,434 |
|  | MPSS Capital | 5,691 | 10,001 | 15,785 | 12,501 | 14,548 |
|  | MPSS Labor | 152 | 133 | 95 | 31 | 10 |
| Cross Sectional 2009 | Capital/Labor | 37.5 | 76.9 | 166.7 | 391.3 | 1,470 |
|  | MPSS Capital | 3,662 | 6,111 | 9,565 | 13,965 | 19,142 |
|  | MPSS Labor | 94 | 80 | 59 | 36 | 13 |

**Appendix C. The flexibility tradeoff between production function and inefficiency distribution assumptions.**

The potential exists for a flexibility tradeoff between modeling the production function nonparametrically and modeling the unobserved inefficiency without distribution assumptions. Since our simulated datasets consist of well-distributed data along the input-output space, we do not need an MBCR prior to obtain computationally efficient runs of MBCR-I. Nevertheless, the more clustered and uneven distribution of observations in our application datasets benefit computationally from a more informative prior on the curvature of the function given by MBCR. When we run MBCR on both datasets, the number of hyperplanes needed to describe the production frontier decreases when inefficiency is introduced. The results for MBCR-I in Table C1 and Table 10 are the same.

**Table C1. Fitting statistics comparison for MBCR and MBCR-I for the Japanese concrete industry Dataset.**

| MBCR | # Hyperplanes fitted | Median Inefficiency | % SS Model | % SS Noise |
|---|---|---|---|---|
| Cross Sectional 2007 | 2.22 | 0% | 55% | 45% |
| Cross Sectional 2008 | 2.18 | 0% | 55% | 45% |
| Cross Sectional 2009 | 3 | 0% | 53% | 47% |
| Cross Sectional 2010 | 4 | 0% | 55% | 45% |
| Panel | 4 | 0% | 56% | 54% |

| MBCR-I | # Hyperplanes fitted | Median Inefficiency | % SS Model | % SS Noise |
|---|---|---|---|---|
| Cross Sectional 2007 | 2.11 | 29% | 66% | 34% |
| Cross Sectional 2008 | 2.03 | 28% | 75% | 25% |
| Cross Sectional 2009 | 2.60 | 29% | 70% | 30% |
| Cross Sectional 2010 | 2.36 | 27% | 76% | 24% |
| Panel | 3.63 | 43% | 81% | 19% |

Note that an omitted inefficiency model results in a more complex nonparametric shape-constrained frontier estimation, whereas the use of a moderately flexible inefficiency model results in frontier estimations with fewer hyperplanes. We believe very general inefficiency distribution specifications, especially non-parametric specifications considering non-monotonic or heteroscedastic behaviors, will result in coarser production frontier estimates. That is, if both the frontier specification and the estimated values are very flexible, we may encounter an identification problem. We emphasize that this insight is only possible due to the one-stage nature of MBCR-I.

**Appendix D. A coarse grid search algorithm to compute MPSS on a multivariate setting.**

Calculate MPSS at a given Capital/Labor ratio as follows:

1. Let $\boldsymbol{\omega} = (\omega_1, \ldots, \omega_L)$ be a uniform grid of $L$ points across the Labor axis.
2. Set $R = Capital/Labor$ to a value of interest; for example, a percentile of the empirical Capital/Labor distribution.
3. Evaluate the predicted frontier output value $\hat{f}_\ell(R \cdot \omega_\ell, \omega_\ell)$ for each $\ell \in \{1, \ldots L\}$.
4. Obtain $\beta_{[R \cdot \omega_\ell, \omega_\ell]}$, the regression coefficients of the hyperplane supporting the given $(R \cdot \omega_\ell, \omega_\ell)$ pair.
5. Define a measure of aggregate input $I = \beta_1 R \omega_\ell + \beta_2 \omega_\ell$.
6. Calculate $MPSS^* = \max_{\ell \in \{1, \ldots, L\}} \hat{f}(R \cdot \omega_\ell, \omega_\ell) / I$.